\documentclass{WileyMSP-template}

\usepackage{lineno}
\usepackage{amsmath}
\usepackage{amssymb}
\usepackage{textcomp}  
\usepackage{gensymb} 
\usepackage{bm}

\begin{document}

\pagestyle{fancy}
\rhead{\includegraphics[width=2.5cm]{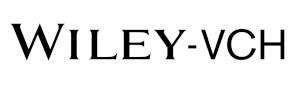}}

\title{Tailoring ultra-high-order optical skyrmions}
\maketitle


\author{Xinji Zeng$\And$, Jing Fang$\And$, Haijun Wu, Jinwen Wang, Yun Chen, Yongkun Zhou, Xin Yang, Chengyuan Wang, Dong Wei, Haixia Chen, Hong Gao*, and Yijie Shen$\dagger$}


\dedication{$\And$The authors contributed equally to
this work.}

\begin{affiliations}
X. Zeng, J. Wang, Y. Zhou, X. Yang, C. Wang, D. Wei, H. Chen, H. Gao \\
Ministry of Education Key Laboratory for Nonequilibrium Synthesis and Modulation of Condensed Matter, Shaanxi Province Key Laboratory of Quantum Information and Quantum Optoelectronic Devices, School of Physics, Xi'an Jiaotong University, Xi'an 710049, China\\
E-mail: honggao@xjtu.edu.cn

J. Fang, H. Wu, Y. Shen\\
Center for Disruptive Photonic Technologies, School of Physical and Mathematical Sciences $\And$ The Photonics Institute, Nanyang Technological University, Singapore 637378, Singapore\\
E-mail: Yijie.shen@ntu.edu.sg\\
Y. Chen\\
Department of Physics, Huzhou University, Huzhou 313000, China\\
Y. Shen\\
Department of Electrical and Computer Engineering, National University of Singapore, Singapore, 117583, Singapore
\end{affiliations}



\begin{abstract}

Skyrmions, as quasiparticles with topological spin textures, has recently garnered great attention for both condensed matter and structured wave communities, promising next-generation large-density robust information technologies. However, a big challenge to this end is that the generation of high-order skyrmions is elusive in any physical systems. Here, we propose the method to create and control ultra-high-order skyrmions (skyrmion number up to $400^{th}$) in a structured light system. We also experimentally control the topological state transition between bimeron and skyrmion, arbitrarily tailor the transverse size of an arbitrary-order skyrmionic beam independent of topological number, and ensure the topological stability upon propagation.  Our work offers solutions for topologically resilient communication and memory with much enhanced information capacity.

\end{abstract}


\section{Introduction}
Skyrmions, as topological spin textures, initially proposed by British particle physicist Tony Skyrme to describe the localization of meson fields, are quasiparticle entities with or without effective mass~\cite{skyrme1962unified}. The nontrivial topological structure of skyrmions, which directly affects the static and dynamic behavior of magnetic domains, has led to their extensive study in condensed matter physics~\cite{mühlbauer2009skyrmion,yu2010real,bogdanov2020physical,tai2024field}. Magnetic skyrmions excited by out-of-plane magnetic fields at material surfaces can be extremely small and driven by weak currents, holding promise for next-generation high-density information storage and logic manipulation devices~\cite{schwarze2015universal,fert2017magnetic,seki2020propagation}. Since topological types and topological numbers are the main information carriers for skyrmions, it is urgent to construct diverse and higher-order quasiparticles. For example, merons with fractional topological charges~\cite{yu2018transformation}, bimerons as homotopy classes of skyrmions~\cite{li2020bimeron}, skyrmion bags~\cite{foster2019two} and skyrmion bundles~\cite{tang2021magnetic} reported in recent years. In the fields of nonlinear optics~\cite{karnieli2021emulating} and condensed matter~\cite{shustin2023higher}, efforts to generate high-order skyrmions have been made. However, for example, magnetic materials tend to favor lower energy states, making the generation of higher-order skyrmions extremely difficult in any physical system.

Optical skyrmions are initially generated by disrupting the transverse nature of electromagnetic fields within evanescent waves, encompassing both field and spin skyrmions~\cite{tsesses2018optical,du2019deep,shen2024optical}. The polarization of light, which embodies its wave nature, directly correlates with the spin of photons. Consequently, skyrmions can be created in free space through suitable polarization configurations, for instance, tight focusing effects~\cite{zeng2024tightly} and Stokes vectors~\cite{shen2022generation,mcwilliam2023topological,shen2021topological,cisowski2023building} are used, the latter is also known as optical paraxial skyrmionic beams or Stokes skyrmions. Stokes skyrmions can be generated inside\cite{yoneda2024green} or outside\cite{tamura2024direct,teng2023physical} a laser cavity, and even in integrable on-chip devices\cite{lin2024chip}, link the skyrmion number with orbital angular momentum (OAM), offering potential for optical communications~\cite{berkhout2010efficient,PhysRevLett.92.013601,karimi2009efficient,wan2023ultra,RNoamErasure}, quantum entanglement source preparation\cite{ornelas2024non} and quantum storage~\cite{dong2023highly}. More importantly, OAM is theoretically infinite-dimensional in Hilbert space, making it a good candidate for generating higher-order skyrmions. Beyond the typical Laguerre-Gaussian (LG) beams, perfect vortex (PV) beams also carry OAM~\cite{ostrovsky2013generation,vaity2015perfect,pinnell2019perfect,shen2019optical}. The size of PV beams do not increase with increasing OAM, which is distinctly different from LG modes, making them more favored in light-matter interaction applications~\cite{zeng2024spatial,forbes2021structured} such as particle manipulation~\cite{chen2013dynamics,wang2024topological} and optical memory~\cite{chen2021phase,zeng2023optical}.

In this paper, we explore the potential of constructing ring-shaped optical quasiparticles using topological methods, and experimentally tailor ultra-high-order skyrmionic beams through the coherent superposition of typical or perfect modes. Such high-order skyrmions have never been generated in any physical systems. We not only generate the highest 400th-order skyrmions that significantly boost information capacity but also introduce a scheme for controlling the transverse beam size. The application of magnetic quasiparticles in high-density data storage~\cite{han2022high,tomasello2014strategy} gives us confidence to anticipate the future use of optical quasiparticles in optical memory and even the construction of topologically enhanced classical and quantum information networks. 

\section{Concepts}

 \begin{figure*}[ht!]
    \vspace*{0mm} 
        \centering
        \includegraphics[height=9cm]{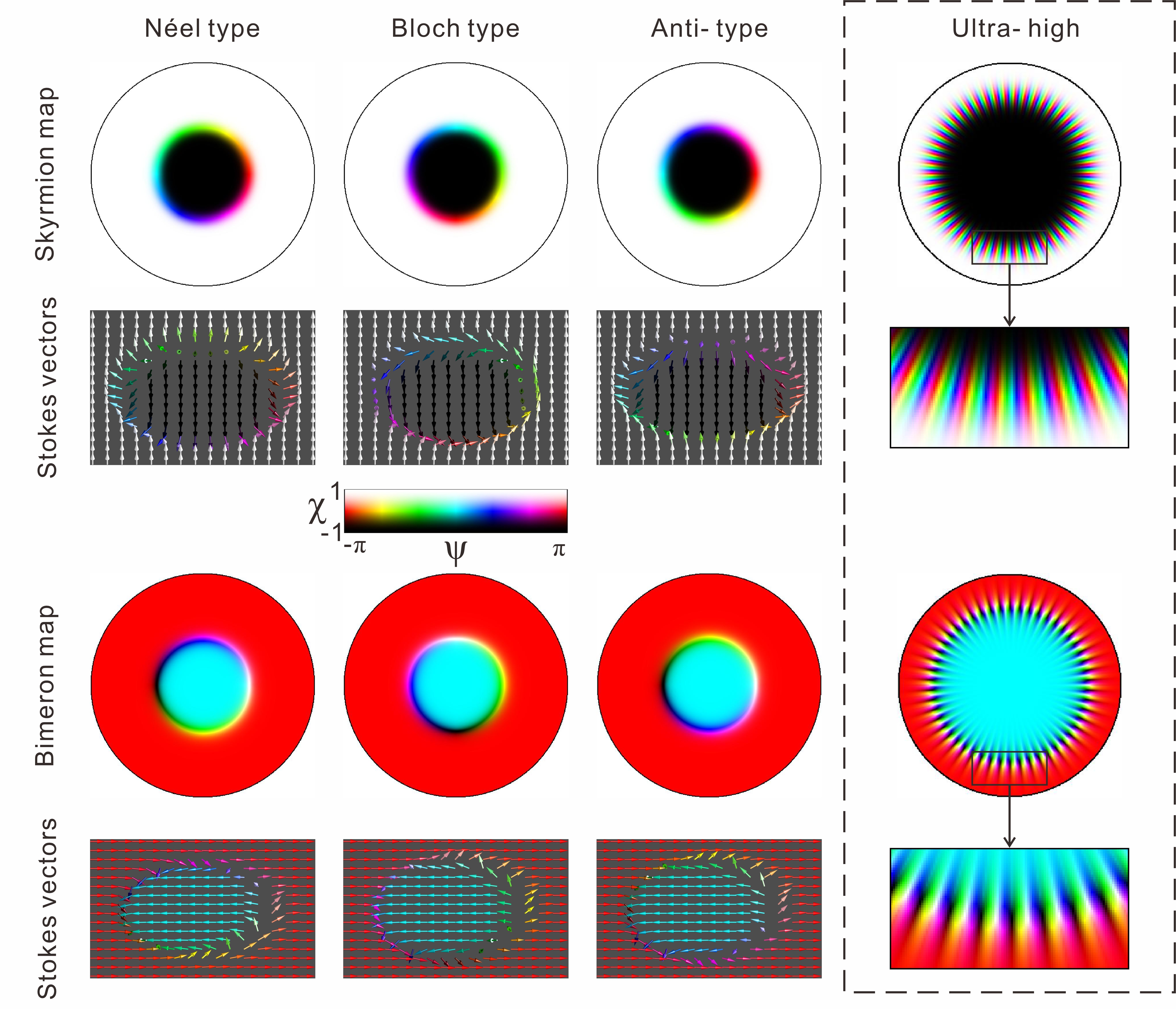}
        \\[0mm] 
        \caption{Theoretical concept of ring-shaped skyrmions and bimerons, with columns from left to right representing N\'eel, Bloch, and Anti-types.
 Top row, skyrmion maps and stokes vectors, the central black and peripheral white represent left - and right - handed circular polarization respectively; bottom row, the bimeron maps and stokes vectors, the central blue and peripheral red represent vertical and horizontal linear polarization respectively. Colorbar is based on the Poincar\'e sphere's azimuth angle $\Psi$ and spatial angle $\chi$. The right column presents the maps of ultra-high-order skyrmions and bimerons along with their local magnified images. The main feature of the ring-shaped quasiparticles is that the center is no longer point-like but disk-like. The simulations are all synthesized by Laguerre-Gaussian beams with opposite unequal-order orbital angular momentum.
} 
        \label{fig1}
    \end{figure*}
    
In mathematics, a 2-dimensional (2D) sphere in a real number space is topologically equivalent to the union of a 2D real number plane with the point at infinity:
\begin{equation}\label{eq1}
S^{2}=\{(x,y,z)\in \mathbb{R}^{3}\mid x^2+y^2+z^2=1\}\cong \mathbb{R}^{2}\cup\{\infty\},
\end{equation}
which means there should always be a continuous bijective mapping between these two topological spaces. Considering the South Pole of the unit sphere tangent to the 2D real plane, the following stereographic projection can be constructed:
\begin{equation}\label{eq2}
f:(x,y,0)\mapsto (\frac{4x}{x^2+y^2+4},\frac{4y}{x^2+y^2+4},\frac{2(x^2+y^2)}{x^2+y^2+4}),
\end{equation}
it's evident that $f$ is invertible, and both $f$ and $f^{-1}$ are continuous mappings. As the points on the plane $(x,y,0)$ extend towards infinity, $x^2+y^2$ being an order of magnitude greater than $x$ and $y$, the corresponding points on the sphere approach coordinates of $0$ for the first two dimensions, while the z coordinate $2-8/(x^2+y^2+4)$ converges to $2$, which is indeed the North Pole of the sphere. Another unique point is the tangency between the sphere and the 2D real plane at $O=(0,0,0)$, which corresponds to both the South Pole of the sphere and the origin of the plane. If the two topological spaces in Eq.\eqref{eq1} are parameterized, for paraxial optics, such as the Stokes vectors, then the aforementioned mapping process corresponds precisely to the construction of a 2D skyrmionic structure ('baby skyrmions'\cite{shen2024optical}), as shown in Fig.\ref{fig1}.

Optical Skyrmions are typically described as coherent combinations of orthogonally polarized vortex and non-vortex beam, exemplified by LG beams:
\begin{equation}~\label{eq3}
    \bm{E_s}(r,\phi)=\cos(\theta)\mathrm{LG}^{l_1}_0(r,\phi)\bm{\mathrm{e_r}}+\sin(\theta)\mathrm{LG}^{l_2}_0(r,\phi)\exp(i\Phi)\bm{\mathrm{e_l}},
\end{equation}
where $\mathrm{LG}^l_0(r,\phi)=\sqrt{2/(\pi|l|!)}(\sqrt{2}r/w_0)^{|l|}\exp(-r^2/w_0^2)\exp(il\phi)$ is the Laguerre-Gaussian function, $w_0$ is the beam waist, $|l_2|>l_1=0$ and $\theta\in[0,\pi/2]$ represents the relative amplitude of the two modes, $\Phi$ denotes the relative phase, and $\bm{\mathrm{e_r}},\bm{\mathrm{e_l}}$ represent the unit vectors for  right circularly polarized (RCP) and left circularly polarized (LCP) components, respectively. Such approach is effective for generating low-order skyrmionic beams, but it becomes quite different for higher-order cases. To illustrate this, we first need to understand that for the 3D Stokes vector $\bm{S}$, a topological invariant known as the winding number (or topological charge) can be defined as\cite{liu2016skyrmions}:
\begin{equation}\label{eq4}
    N=\frac{1}{4\pi}\iint \bm{S}\cdot(\frac{\partial\bm{S}}{\partial x}\times\frac{\partial\bm{S}}{\partial y})\,\mathrm{d} \bm{r}^2,
\end{equation}
in polar and spherical coordinates, $\bm{r}=(r\cos \phi, r\sin \phi)$ and $\bm{S}=(\cos \Phi(\phi)\sin\theta(r), \sin\Phi(\phi)\sin\theta(r),\\ \cos\theta(r))$, then Eq.\eqref{eq4} is $\frac{1}{4\pi}\int_0^\infty\,\mathrm{d}r\int_0^{2\pi}\,\mathrm{d}\phi\frac{\mathrm{d}\theta(r)}{\mathrm{d}r}\frac{\mathrm{d}\Phi(\phi)}{\mathrm{d}\phi}\sin\theta(r)=\frac{1}{4\pi}\cos\theta(r)\vert_{r=0}^{r=\infty}\Phi(\phi)\vert_0^{2\pi}=m\cdot n$, where $m,n$ is the polarity and vorticity. Clearly, the direction and times of the vectors covering the 3D parameter sphere determine the magnitude of $N$. Experimentally, to achieve a well-defined $N$ and considering the spatial intensity distribution of the LG beam, the two modes in Eq.\eqref{eq3} should have appropriate radial overlap. The spot size of LG beams is proportional to their order, hence higher-order ($|l|>10$ here) LG beams will not have much effective overlap with the 0th-order mode, suggests an undesirable winding number.

To effectively generate higher-order skyrmions, $l_1$ in Eq.\eqref{eq3} need not be 0 but should satisfy $|l_2|>|l_1|$. In this case, 'baby skyrmions' transform from disk-shaped to ring-shaped. We proceed with some intriguing topological manipulations to illustrate this idea: remove the South Pole of the parameter sphere and excise a circular disk from the 2D real plane centered at the origin. Define the remaining part of $\mathbb{R}^{2}$ as $X=\mathbb{R}^{2}\backslash D$, where $D$ is an open disk: \begin{equation}\label{eq5}
    D=\{x\in \mathbb{R}^{2}\mid \Vert x-O\Vert<r \},
\end{equation}
where $r$ is the radius of the disk. An equivalence relation can be defined on $X$: for $\forall (x,y)\in X$, if both $x$ and $y$ are on the boundary $\partial D$ of $D$, then $x\sim y$; if only $x$ is on $\partial D$, then $x\sim O$; otherwise, $x\sim x$. Thus, for the quotient space $Y=X/ \sim$, all points on $\partial D$ are equivalent to $O$, leading to the following quotient mapping:\begin{equation}\label{eq6}
    p:X \rightarrow Y,x\mapsto \langle x\rangle,
\end{equation}
for $x\in \partial D$, $p(x)=O$; the rest $x\notin \partial D$ are mapped to their respective equivalence classes $\langle x\rangle$. Next, we need to define a surjection $q$: for $\forall x\in X$, if $x\notin \partial D$, then $q(\langle x\rangle)=x$; if $x\in \partial D$, then $q(\langle x\rangle)=O$. It is evident that this surjection is invertible and continuous, effectively showing that $Y \cong \mathbb{R}^{2}$. Through the mapping $p\circ q$, which compresses the boundary $\partial D$, we actually glue $X$ and $O$ into $\mathbb{R}^{2}$. In the Stokes parameter space, $O$ represents a downward-pointing vector. Above mathematical process shows this vector's point distribution is equivalent to a disk-like arrangement, thus achieving the simple idea of constructing a ring-shaped skyrmionic structure.

 The top row of Fig.\ref{fig1} presents the map and Stokes vectors distributions for three types of ring-shaped 1st-order skyrmions. The central (boundary) vector directions of the three quasiparticles are uniformly downward (upward), with differences lying in the color transitions: the first column has a hedgehog-like structure, which is of the N\'eel type; the second column is defined as Bloch type, featuring a vortex structure; the last column is saddle-shaped, with the opposite topological charge compared to the first two, and is of the Anti-type. The bottom row of Fig.\ref{fig1} displays the ring-shaped bimeron structures, which are isomorphic to skyrmions and correspond to the three aforementioned types. It can be observed that each type of bimerons is composed of a pair of merons with the same $N=1/2$ but opposite signs, similar to vortices pair before the topological phase transition occurs~\cite{kosterlitz2018ordering}.

 \section{Experimental setup}

  \begin{figure}[ht!]
    \vspace*{0mm} 
        \centering
        \includegraphics[height=6cm]{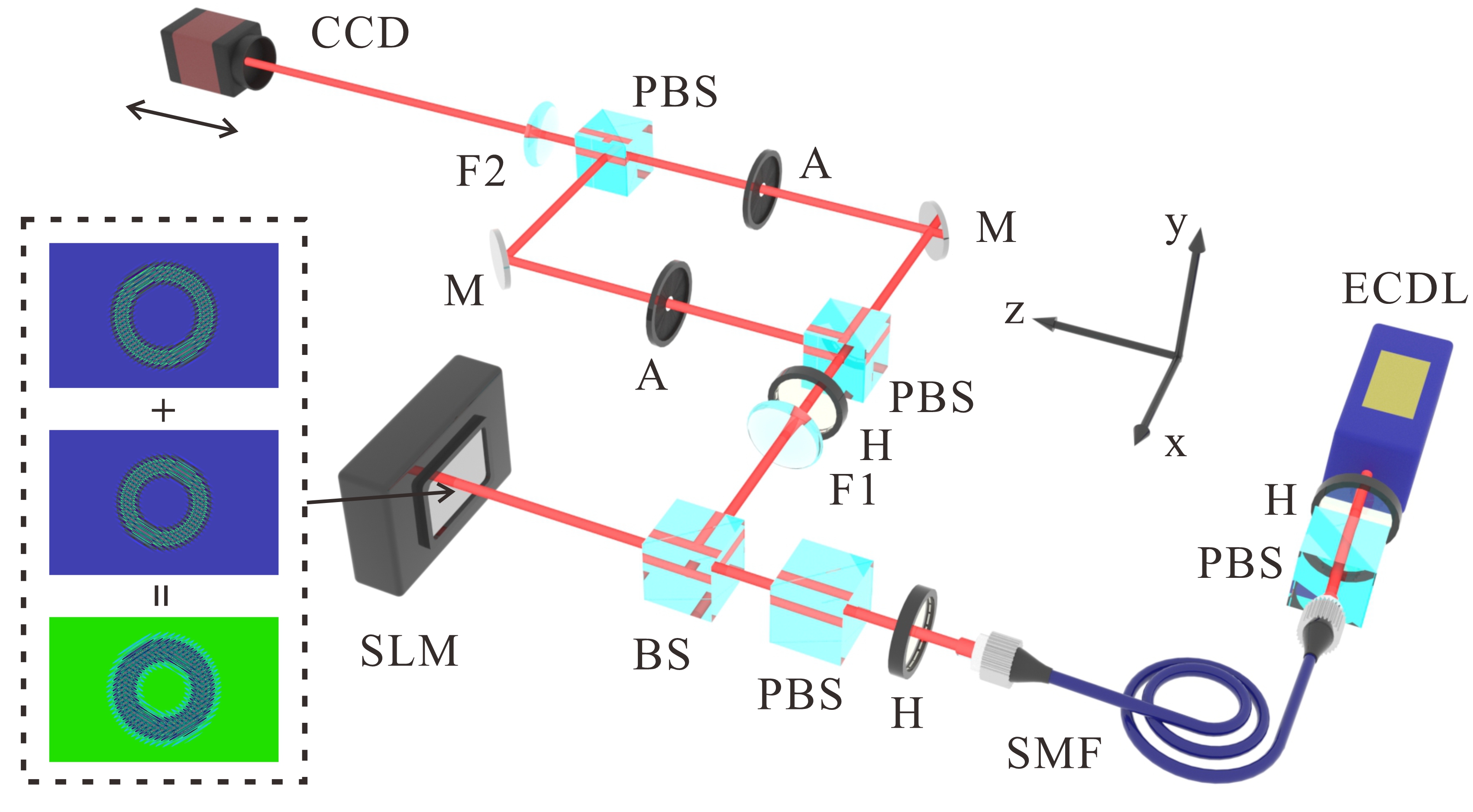}
        \\[0mm] 
        \caption{Experimental setup, ECDL: external cavity diode laser, H: half-wave plate, PBS: polarization beam splitter, SMF: single mode fiber, BS: beam splitter, SLM: spatial light modulator, F1,F2: lens with $f=300$ mm, M: mirror, A: aperture, CCD: movable charge-coupled camera, inset is the multiplexing grating. The Stokes parameters are measured using a polarizing analysis setup (not shown here) consisting of a HWP, a quarter-wave plate (QWP), and a PBS.} 
        \label{fig2}
    \end{figure}
    
We generate the desired beam by complex amplitude-modulating a spatial light modulator (SLM) with multiplexed gratings. As shown in Fig.\ref{fig2}, an external cavity diode laser (ECDL) emits a horizontally polarized beam at 795 nm, which is coupled into a single-mode fiber (SMF) to improve the spatial mode. Due to the SLM's ability to modulate only horizontally polarized light, the combination of a HWP and a PBS is employed prior to the beam's incidence on the SLM. Additionally, a $50:50$ beam splitter (BS) serves to divide the unmodulated and modulated light. 

The SLM adjusts only the phase of the light, but a particular phase distribution can mimic complex amplitude modulation\cite{rosales2017shape}, as depicted in Fig.\ref{fig2}'s inset, where two orthogonally aligned blazed gratings produce a multiplexed grating through point multiplication. To capture the complex amplitude-modulated beam from the SLM, a 4f imaging system is required between the SLM and the charge-coupled camera (CCD). A Mach-Zehnder (MZ) interferometer is positioned between the two lenses. The modulated horizontally polarized beam first passes through a HWP oriented at $\pi/8$ to the horizontal, converting it into diagonally polarized light. The PBS then splits the modulated light into two beams of equal intensity, directing them to the two arms of the interferometer. Each arm has a spatial filter inserted to select the $+1$ order diffracted beams from each grating on the SLM. The output from the MZ interferometer represents the two orthogonally polarized modes described by Eq.\eqref{eq3}. By precisely adjusting the mirrors and PBSs, the two modes are coherently combined to generate the desired beam, which is then detected by a movable CCD.
 \begin{figure*}[hb!]
    \vspace*{0mm} 
        \centering
        \includegraphics[height=7cm]{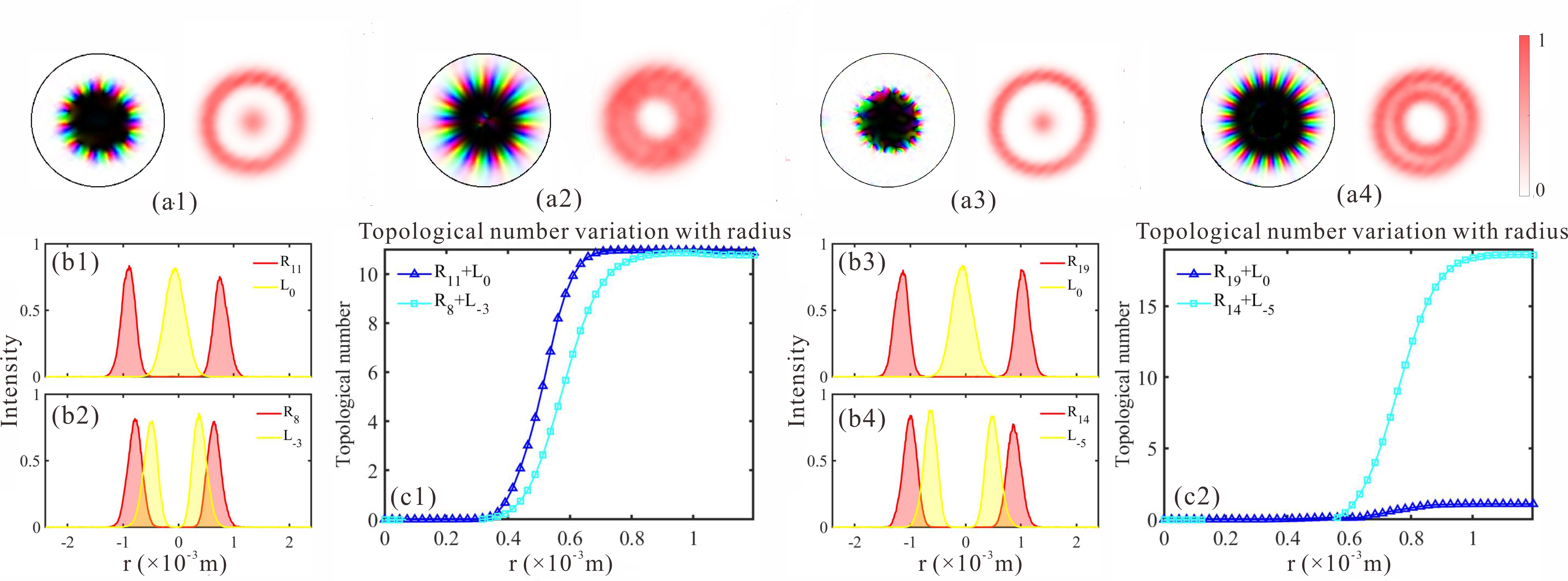}
        \\[0mm] 
        \caption{Comparison of experimental results of disk-shaped and ring-shaped high order skyrmions. The left is 11th-order skyrmions, \textbf{(a1), (a2)}, map and intensity distribution of superimposed modes: $R_{11}+L_0$ and $R_{8}+L_{-3}$; \textbf{(b1), (b2)}, radial intensity curves of the two superimposed modes; \textbf{(c1)}, topological number variation with integration radius ($\times 10^{-3}$ m). The right is 19th-order skyrmions, captions for the corresponding letters are the same as aforementioned, two superimposed modes are $R_{19}+L_{0}$ and $R_{14}+L_{-5}$. $R_l$ and $L_l$ is the lth RCP and LCP LG mode, respectively. The beam waist $w=0.3$ mm.},  
        \label{fig3}
    \end{figure*}

\section{Results and discussions}
\subsection{Nonzero-unequal orders LG beams as basic modes}
Here we show the main experimental results of ultra-high-order skyrmions. Fig.\ref{fig3} presents the generated distinct disk and ring-shaped skyrmions of 11th and 19th-order using LG modes. After applying Eq.\eqref{eq4} and performing some noise reduction on the images, for typical method, the calculated winding number of the two high-order skyrmions are $N=10.7636, 0.9975$, respectively. As is shown in Fig.\ref{fig3}(a3) and (b3), the 19th-order modes overlap minimally with the 0th one in terms of intensity distribution, thus the skyrmion map transitioning chaotically from the dark center to the bright periphery. The red curve in Fig.\ref{fig3}(c2), which shows the topological number variation with integration radius, confirms this observation. Although the 11th-order skyrmion maintains a robust topological structure, the 19th-order case significantly deviates from expectations. But for the ring-shaped case the calculated topological number are $N=10.7636, 18.8301$, which confirmed the efficacy in generating higher-order skyrmions. But unfortunately, when $|l|$ and $r$ are large, the $\mathrm{LG}^l_0(r,\phi)$ itself will lead to a clash: $r^{|l|}\to\infty,\,\exp(-r^2)\to0$. Here we prevent this by approximating with $\mathrm{LG}^l_0(r,\phi)\approx \exp-((r-w_0\sqrt{|l|/2})/w_0/\sqrt{2})^2\exp(il\phi)$. 
 \begin{figure}[ht!]
    \vspace*{0mm} 
        \centering
        \includegraphics[height=11cm]{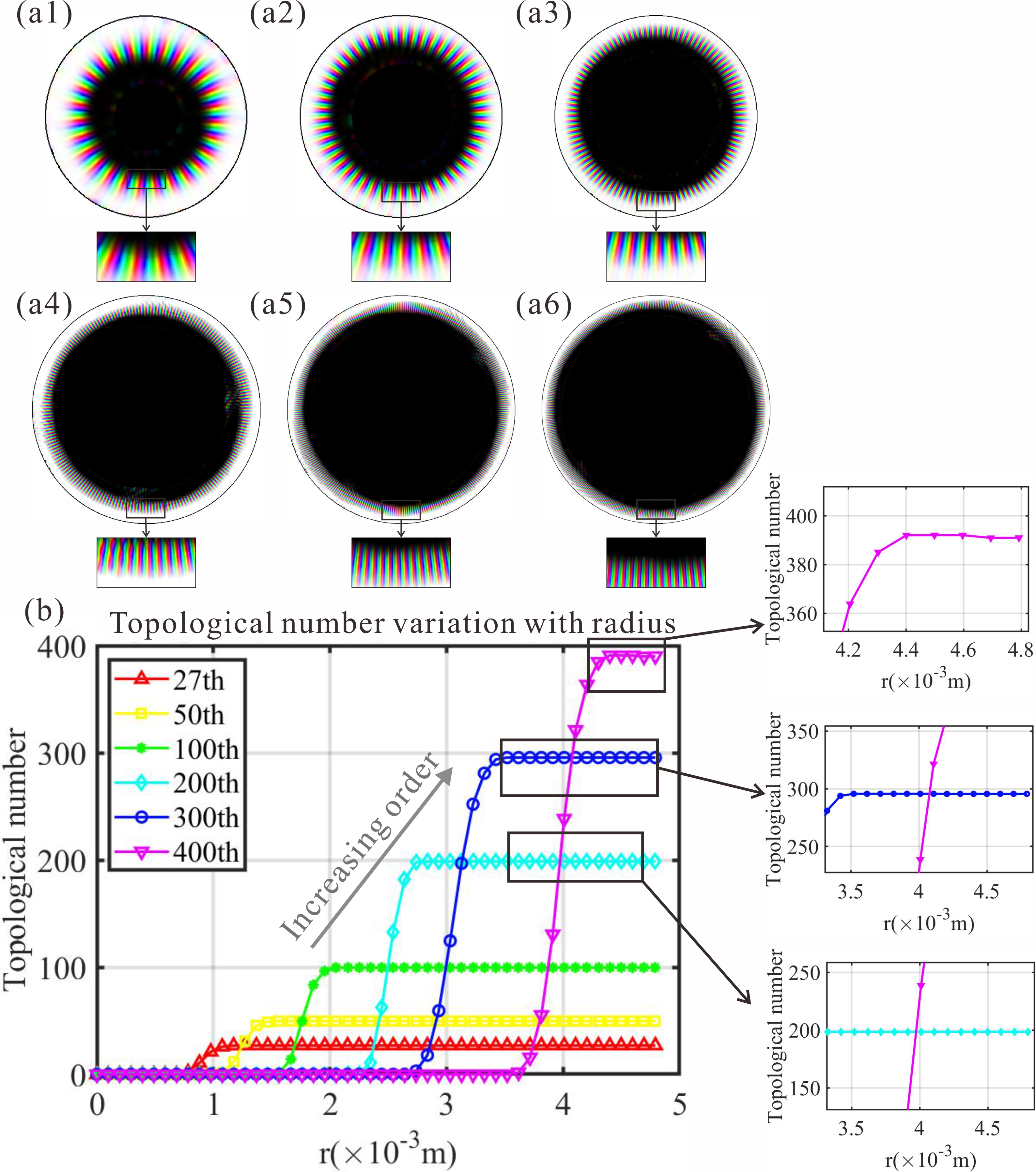}
        \\[0mm] 
        \caption{Experimental results of ultra-high-order skyrmions. \textbf{(a1)-(a6)} is 27th, 50th, 100th, 200th, 300th and 400th-order skyrmion with each local magnified image, respectively; \textbf{(b)} is topological number variation with integration radius ($\times 10^{-3}$ m) of each ultra-high-order skyrmion with local magnified plots, the gray arrow indicates the trend of curve changes. The beam waist is the same as Fig.\ref{fig3}.} 
        \label{fig4}
    \end{figure}

Fig.\ref{fig4} presents the results of generated ultra-high-order skyrmions. The calculated winding numbers are $N=26.7343, 49.7268, 99.8110, 199.291, 296.5260, 392.0620$ for the 27th, 50th, 100th, 200th, 300th and 400th-order skyrmions. 
From the magnified view of the skyrmion map, it is evident that the angular periodicity increases with the order, implying the Stokes vectors wrap around the parameter sphere along the latitude more times. As the topological number increases, the number of periodic changes in the azimuthal polarization increases, making the generated ultra-high-order skyrmion appear to have only some moir\'e fringes. From the curves of the topological numbers of the skyrmions of different orders versus integration shown in Fig.\ref{fig4}(b), it can be seen that the radial distance occupied by the polarization transition from the black center to the white edge is the same, and the slopes of the corresponding curves are not significantly different. This indicates that the method we used can ensure the effective overlap of the two orthogonal modes. Additionally, the overall shift of the curve to the right indicates that the size of the generated quasiparticles increases with the order. It should be noted that weak light intensities detected by the CCD in experiments can introduce noise, but proper data processing can mitigate its impact on the results (referring to color maps, with negligible effects on topological numbers), and during the polarization projection measurement, we fine-tuned the position of the PBS to achieve the optimal extinction ratio. 

Although we have confirmed that superimposing vortex beams of unequal order with orthogonal polarizations can yield good topological structures, there's still an upper limit to the order of skyrmions we can produce experimentally\cite{pinnell2020probing}. Consider the following aspects. 
(1): Physically, high OAM density can lead to the generation of evanescent waves, hence for an optical system with a numerical aperture of NA, the topological number must satisfies\cite{pinnell2020probing}:
\begin{equation}\label{eq7}
    |N|\leqslant k\,\mathrm{NA}\,R,
\end{equation}
where $k$ is the wave vector, $R$ is the spot radius. 
(2): Experimentally, the screen size and resolution of the SLM and CCD are finite. Constrained by the characteristic that the LG mode's spot size increases with order, the SLM screen must be capable of modulating larger spots, and the CCD screen should also receive spots of corresponding size (diffraction must be taken into account). Moreover, due to the calculation of topological charge based on pixel integration, higher order skyrmions do not have enough pixels for reliable results, so the precision of calculations is particularly crucial. However, existing literature\cite{ruffato2019multiplication} on multiplying OAM by loading specific phases onto input beams through diffraction is expected to raise the upper limit of the quasiparticle order that can be generated. 

Under the dual constraints of physical and device limitations mentioned above, Fig.\ref{fig4}(b) shows the topological charge varies with the integration radius. It's clear that higher orders correspond to larger radial coordinates for reaching maximum charge. Note that the maximum integration radius here is under $5$ mm, due to our SLM resolution of 1900$\times$1200 with a single pixel width of $8\,\mu$m. Although such ultra-high-order topologies can greatly boost information capacity, the large spot is not favored in applications like optical storage. One might consider reducing the waist size to enhance practicality, but this confines the method to digital devices like SLMs, and repeatedly adjusting parameters is undoubtedly counterproductive to practicality. 

\subsection{PV beam: another ring-shaped choice}
    
An effective solution is to generate topological beams using PV beams. Typically, an ideal PV beam possesses arbitrary OAM and a fixed ring width, which can be described by a Dirac delta function: $\delta(r-R)\exp (i l \phi)$, where $R$ is the ring radius and $l$ is OAM number. However, this can't be experimentally realized, a quasi-PV beam with a finite thickness $T$ is an excellent candidate~\cite{pinnell2019perfect,vaity2015perfect}:
\begin{equation}\label{eq8}
    P^l_{R,T}(r,\phi)=i^{l-1}\exp[-\frac{(r-R)^2}{T^2}]\exp(il\phi), R\gg T,
\end{equation}
by coherently combining two PV beams with distinct ring radius and OAM, a perfect topological structure is constructed (here we use perfect bimeronic (PB) beam as an example):
\begin{equation}\label{eq9}
    PB(r,\phi)=\cos(\theta) P^{l_1}_{R_1,T}(r,\phi) \bm{\mathrm{e_x}}+\sin(\theta) P^{l_2}_{R_2,T}(r,\phi)\exp(i\Phi) \bm{\mathrm{e_y}},
\end{equation}
where $l_1\neq l_2\, \&\, R_1\neq R_2$, $\theta$ and $\Phi$ are defined the same as Eq.\eqref{eq3} mentioned above, and $\bm{\mathrm{e_x}},\bm{\mathrm{e_y}}$ represent the unit vectors for horizontal and vertical polarizations, respectively. 

\begin{figure}[ht!]
    \vspace*{0mm} 
        \centering
        \includegraphics[height=7.5cm]{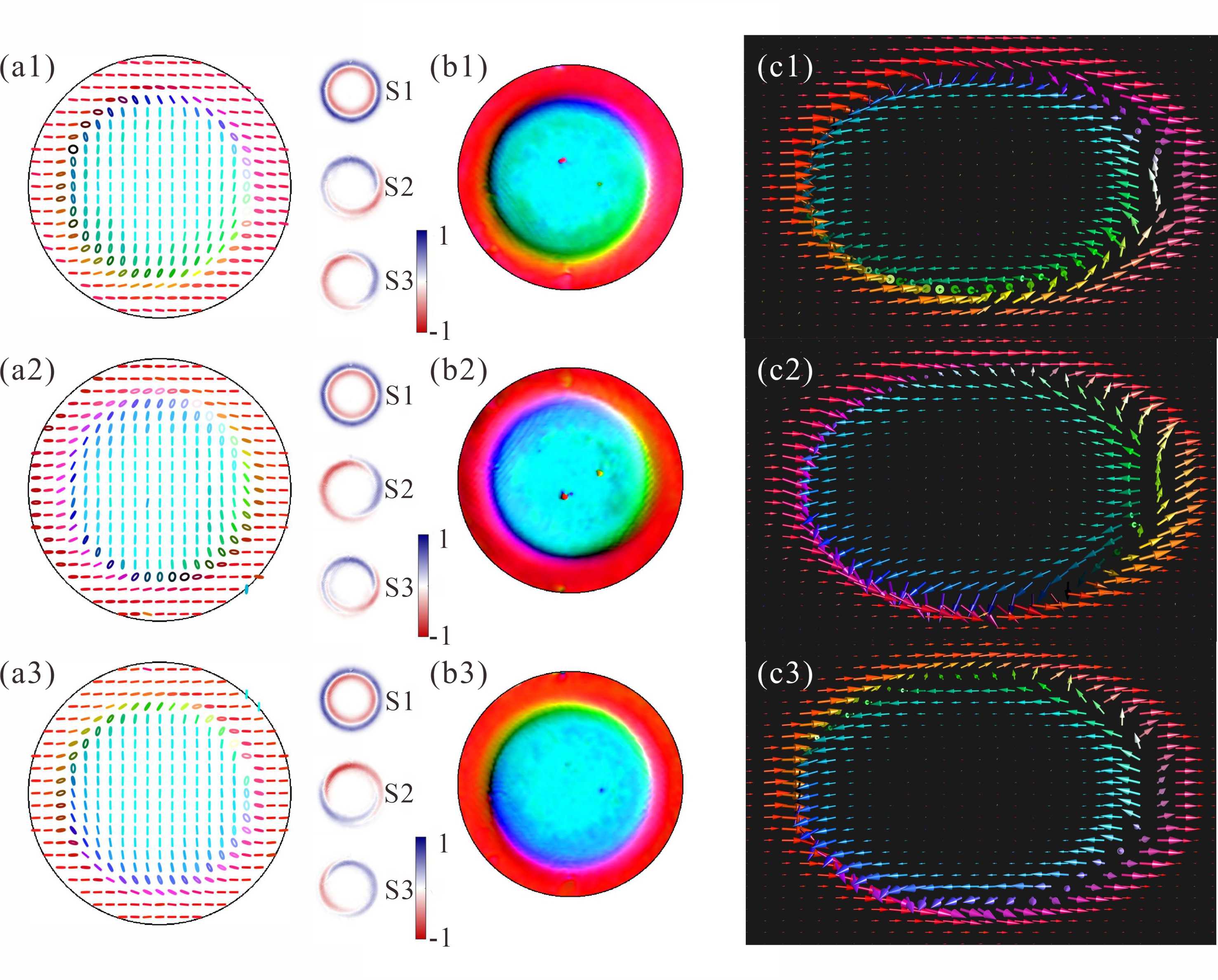}
        \\[0mm] 
        \caption{Experimental results of three different types 1st-order PB beams; The top row is N\'eel type, middle row is Bloch type, and bottom row is Anti-type; \textbf{(a1)-(a3)} are polarization distributions, insets are Stokes parameters; \textbf{(b1-b3)} are bimeronic maps; \textbf{(c1)-(c3)} are Stokes vectors distributions. Parameters used are: $T=0.25$ mm, $R_1=5T,R_2=4T$.} 
        \label{fig5}
    \end{figure}

Fig.\ref{fig5} presents the results of three types of bimerons generated in the experiment. It's clear that the experimental results align well with the simulated outcomes presented in Fig.\ref{fig1}, with minor shifts in the positions of LCP and RCP(dark and bright part in density map) due to the non-strict alignment of the two PV modes, yet it has minimal impact on the outcomes. The calculated topological charges of the three 1st-order bimerons are found to be: $N=0.9923, 0.9893, -0.9923$. The experimental parameters such as $R, T$ are provided in the caption of Fig.\ref{fig5}.

\begin{figure*}[hb!]
    \vspace*{0mm} 
        \centering
        \includegraphics[height=7.5cm]{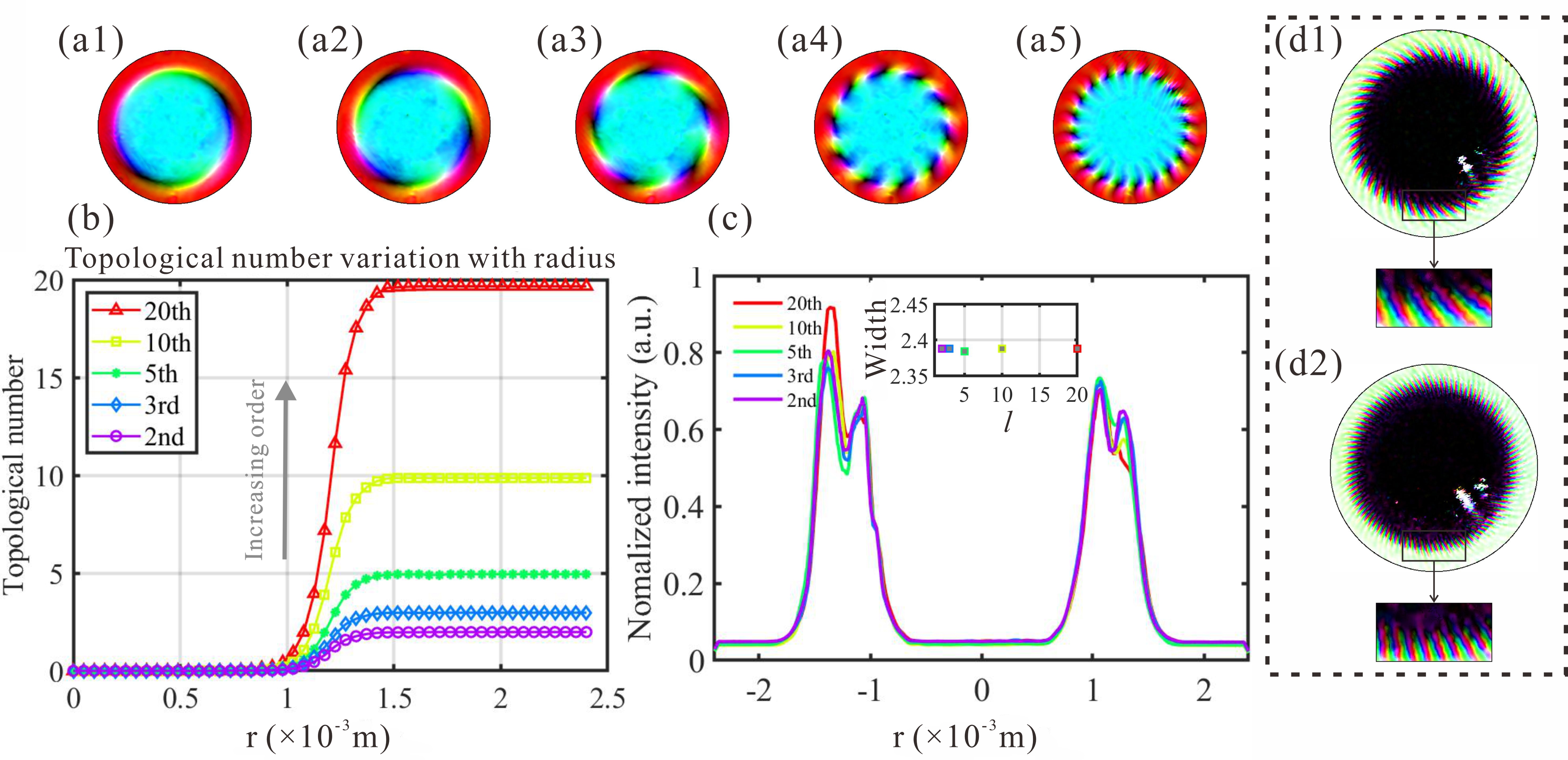}
        \\[0mm] 
        \caption{\textbf{(a1)-(a5)}: Experimental results of 2nd, 3rd, 5th, 10th, 20th-order PB beams, parameters are the same as Fig.\ref{fig5}; \textbf{(b)}, topological number variation with integration radius ($\times 10^{-3}$ m), the gray arrow indicates the trend of curve changes; \textbf{(c)}, radial normalized intensity, inset are ring widths of different orders; \textbf{(d1), (d2)}, results of 50th and 100th order perfect skyrmions and local magnified images, with $T=0.1$ mm, $R_1=17T,R_2=14T$, $l_1=30, 60, l_2=-20, -40$.} 
        \label{fig6}
    \end{figure*}

Given the key feature and advantage of PV beams is that their ring width does not increase with higher OAM values, the experiment successfully generated PB beams of higher orders, as shown in Fig.\ref{fig6}(a1)-(a5). Firstly, the topological charges of the generated bimerons are calculated: $N=1.9830, 2.9765, 4.9425, \\9.8554, 19.6612$ for the 2nd, 3rd, 5th, 10th and 20th-order. Fig.\ref{fig6}(b) presents the variation of the topological number with integration radius. Intuitively, the integration values for PB beams of various orders transit from 0 to $N$ between the center to the edge, indicating the Stokes vectors are well localized within the annular region. Morever, the radial position of the max charge suggests that the PB beams of each order indeed have the same ring width, as more clearly shown in Fig.\ref{fig6}(c). The normalized radial intensity curves of PB beams of different orders essentially overlap. The irregularity of the curves makes it challenging to directly solve the ring width of PB beams, but we can use the difference in radial coordinates at half-peak widths of the two intensity peaks as the ring width. This is illustrated in the inset of Fig.\ref{fig6}(c), as the order increases, the ring width fluctuates by approximately just $0.01$ mm. 

Fig.\ref{fig6}(d1)-(d2) show the experimental results of attempting to generate ultra-high-order perfect skyrmionic (PS) beams, parameters used are provided in the caption. The calculated winding numbers are $N=49.6205, 96.0689$ of 50th and 100th-order PS beams, their ring widths are of course the same. Unlike the LG mode, PV beams' narrower thickness on the SLM leads to weaker target field intensity thus more noises. Besides, high-order Bessel-Gaussian beams as the Fourier transforms of PV beams have wide rings, making spatial filtering challenging. Existing literature~\cite{pinnell2019perfect} has established the relationship between the asymptotic ring width of a single PV beam and OAM, for a PS beam ring width $w(l)$:
\begin{equation}\label{eq10}
    w(l)\propto (\sqrt{T^2(l+1)+2R_1^2}+\sqrt{T^2l+2R_2^2})/2,
\end{equation}
where $R1,R2$ are ring radius of two PV beams in Eq.\eqref{eq9} when $l=0$, $T$ is thickness. Eq.\eqref{eq10} indicates that for larger $l$, $R/T$ should be maximized to maintain a constant ring width, which will inevitably reduce the effective modulation intensity of the SLM. That said, the 100th-order PS beam with a smaller ring width we present here is already sufficient as an ultra-high capacity information carrier, and unlike the inherent limitations of LG modes, these drawbacks may be resolved through technological advances in the future.

    \begin{figure}[ht!]
    \vspace*{0mm} 
        \centering
        \includegraphics[height=12.5cm]{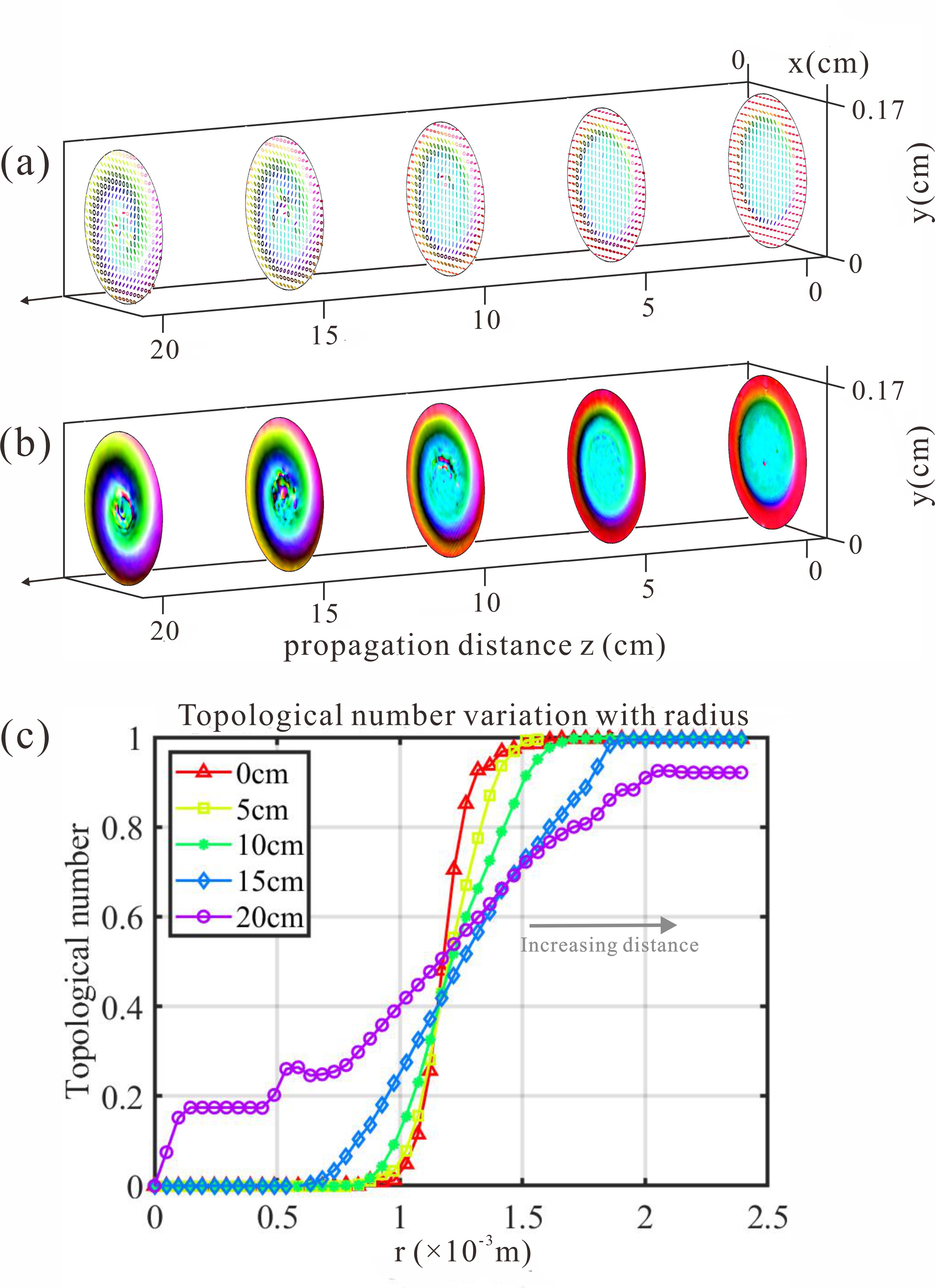}
        \\[0mm] 
        \caption{Propagation properties of 1st-order N\'eel-type PB beam (the top row of Fig.\ref{fig5}) in z direction.
 \textbf{(a)}, changes of polarization during propagation; \textbf{(b)}, changes of bimeronic maps; \textbf{(c)}, topological number variation at different z, the gray arrow indicates the trend of curve changes. Data was collected at positions z=0 cm, 5 cm, 10 cm, 15 cm and 20 cm.
} 
        \label{fig7}
    \end{figure}
    
 Finally, the free-space propagation characteristics of PB beams were investigated. As an integer (ideally), the winding number remains unchanged under continuous deformations of the mapping (corresponding to propagation), thus serving as an invariant constructed for the homotopy groups. To illustrate this, we generally need to construct a continuous mapping $H(s,z)$ to demonstrate the mappings at different $z$ coordinates along the propagation direction are homotopic, that is:
 \begin{equation}\label{eq11}
     C=\{h: S^2 \rightarrow Y\cup\{\infty\}\mid h\,\mathrm{is\,continuous}
\},
 \end{equation}
for $\forall\,h_1,h_2\in C$, there should be a continuous mapping $H:S^2\times [0,1]\rightarrow Y\cup\{\infty\}$ which satisfies:
\begin{equation}\label{eq12}
    H(s,0)\equiv h_1(s),\, H(s,1)\equiv h_2(s),
\end{equation}
where $Y$ has been defined above. Considering the complexity of constructing an explicit form of $H$ mathematically, let's approach this issue from a more qualitative perspective~\cite{beckley2010full}. The complex amplitude of a single PV beam at z is~\cite{vaity2015perfect}:
\begin{equation}\label{eq13}\begin{split}
P_R^l(r,\phi,z)=(-1)^lF(r,z)\mathrm{I}_{l,R}(r,z)\\
\cdot\exp[i\Phi_R(r,z)]\exp[i(\Psi+l\phi+kz)],  
\end{split}
\end{equation}
where $F(r,z)=\frac{T}{Tz}\exp[-\frac{1}{T_z^2}(r^2-\frac{T_z^2}{z_r^2})]$, $T_z=T\sqrt{1+(z/z_r)^2}$ is the thickness at z, $\mathrm{I}_{l,R}(r,z)=\mathrm{I}_l(\frac{2Rre^{i\Psi}}{TT_z})$, $\mathrm{I}_l(\cdot)$ is the $l$th-order modified Bessel function of first kind, $\Phi_R(r,z)=\exp[\frac{ik}{2R_c}(r^2+R^2)]$, $R_c=z+z_r^2/2$, $\Psi=\arctan(z/z_r)$ is the Gouy phase, $z_r=kT^2/2$ is the Rayleigh range. From Eq.\eqref{eq13}, one can see that for the two PV beams in Eq.\eqref{eq9} with different $l,\,R$, the propagation results in only an overall scaling according to $\mathrm{I}_{l,R}$ and a phase difference proportional to $\Phi_R$. The polarization of PB beams merely rotates and twists continuously, which corresponds to the action of $H(s,z)$ on the mappings, thereby proving the topological stability during propagation.

Fig.\ref{fig7}(a)(b) respectively present the polarization and the bimeron map of a 1st-order N\'eel-type PB beam during propagation experimentally. Eq.\eqref{eq13} shows that OAM only affects the phase term, hence it is predictable that higher-order cases will be similar. Intuitively, the polarizations at each z indeed undergo continuous changes of rotation and overall twisting, which aligns with theoretical expectations. To more clearly verify the topological stability during propagation, Fig.\ref{fig7}(c) depicts the relationship between the topological charge and the integration radius at different z. Before propagating to $z=20$ cm, all PB beams effectively achieve $N=1$, but subsequent $N$ values decrease due to excessive noise. The slope in Fig.\ref{fig7}(c) diminishes with distance. While PV patterns may fade over longer distances, a 20 cm propagation distance is adequate for most applications like quantum memory~\cite{parigi2015storage}. Enhanced noise reduction would boost measurement accuracy.

\section{Conclusion}
We have utilized the coherent  superposition of 
various structured beams, including LG or PV modes, to propose an effective method for tailoring ultra-high-order optical skyrmions. In contrast to the conventional way of constructing disk-shaped optical skyrmions by combining vortex and non-vortex beams, we map photonic spin (polarization) textures onto ring-shaped regions without altering their topological properties.

Experiments reveal that the maximum topological charge can reach $400^{th}$ using two nonzero-unequal orders LG beams, which have not been previously reported. Moreover, the generated optical skyrmion using PV modes have a waist size that is independent of topological charge, offering an effective candidate for balancing order and beam width. Lastly, we experimentally confirmed the topological stability of ring-shaped optical skyrmions under continuous deformation using the propagation characteristics of the PV mode as an example. The topological protected ultra-high winding number, as a homotopy invariant, making ring-shaped optical skyrmion as a robust platform for high-capacity, disturbance-resistant classical and quantum communications.

\medskip
\textbf{Acknowledgements} \par 
National Natural Science Foundation of China (No.12404390, 12104358, 12104361, 12304406, 12175168, 92476105 and 92050103); Shaanxi Fundamental Science Research Project for Mathematics and Physics (No. 22JSQ035); Postdoctoral Fellowship Program of China Postdoctoral Science Foundation (No.GZC\\20232118); Shaanxi Province postdoctoral Science Foundation (No.2023BSHEDZZ23) and the Fundamental Research Funds for the Central Universities (No.xzy012023042). Singapore Ministry of Education (MOE) AcRF Tier 1 grant (RG157/23), MoE AcRF Tier 1 Thematic grant (RT11/23), and Imperial-Nanyang Technological University Collaboration Fund (INCF-2024-007), Nanyang Technological University Start Up Grant.

\medskip
\textbf{Conflict of Interest}\par The authors declare no conflicts of interest.

\medskip

%


\medskip
\textbf{Appendix}
\begin{figure}[ht!]
    \vspace*{0mm} 
        \centering
        \includegraphics[height=7cm]{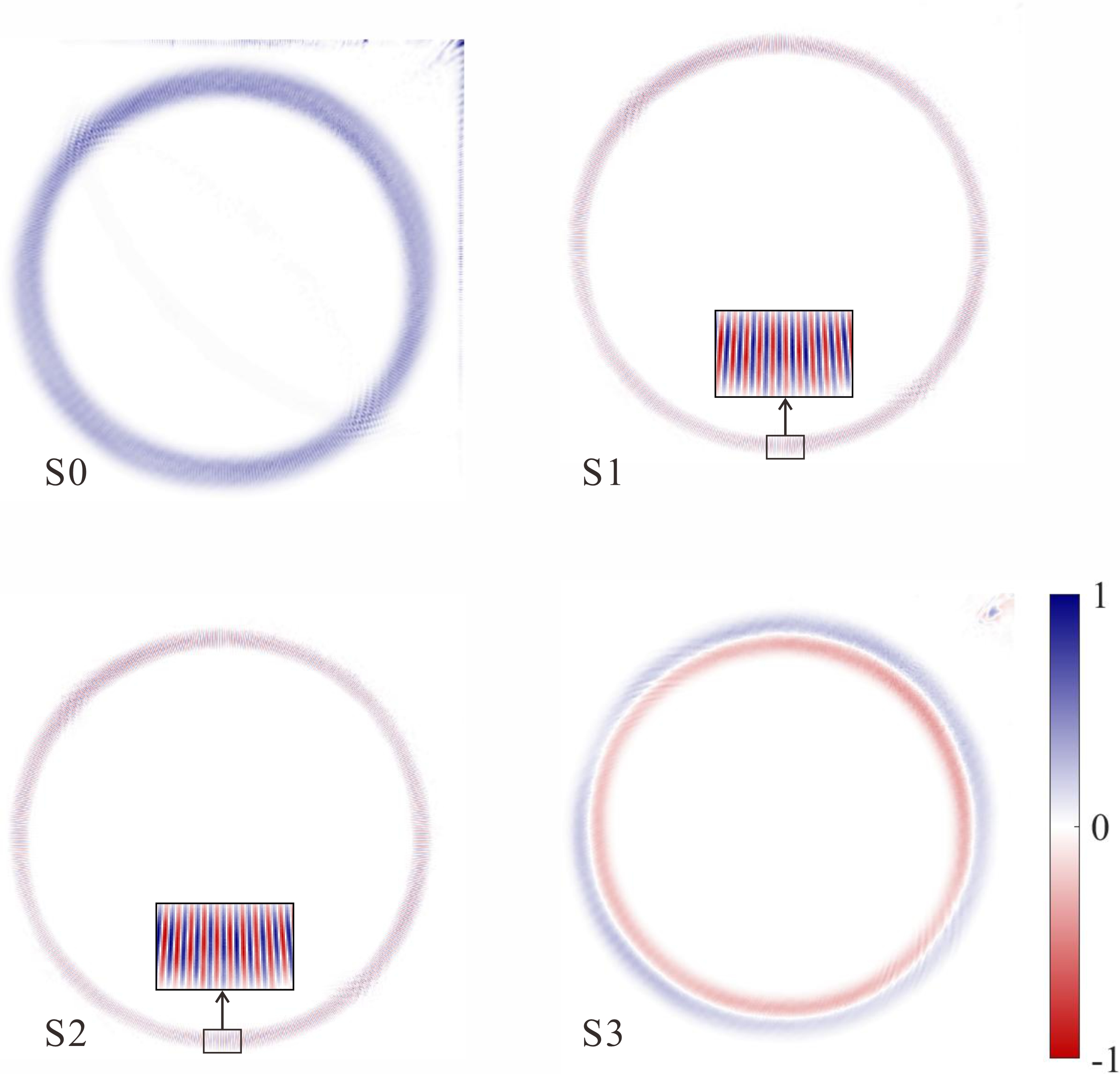}
        \\[0mm] 
        \caption{The experimental Stokes parameters $S0-S3$ of the 400th-order skyrmion generated by LG beams, along with local magnified images of $S1$ and $S2$.} 
        \label{A1}
    \end{figure}
\par
\begin{figure}[ht!]
    \vspace*{0mm} 
        \centering
        \includegraphics[height=7cm]{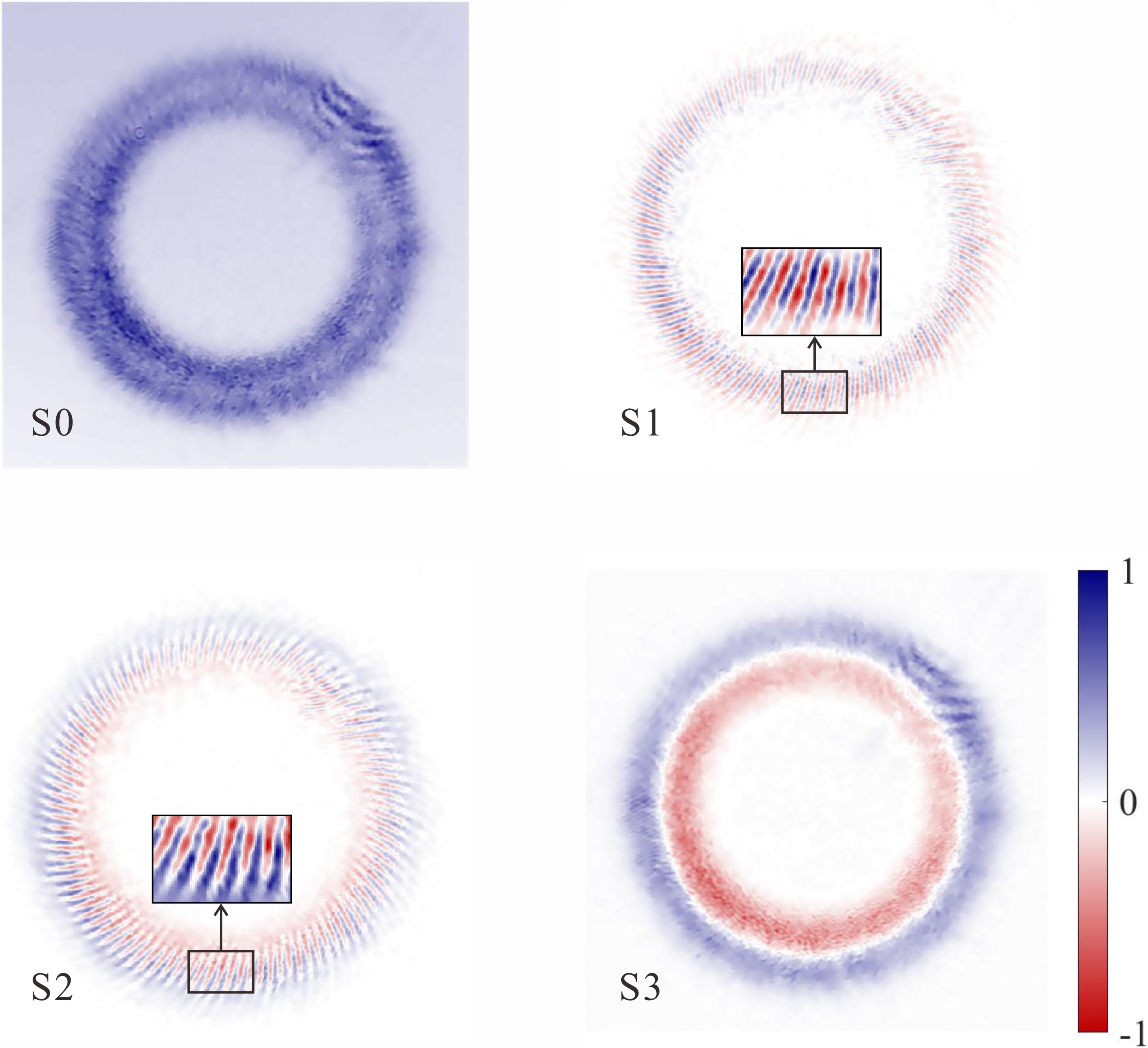}
        \\[0mm] 
        \caption{The experimental Stokes parameters $S0-S3$ of the 100th-order skyrmion generated by PV beams, along with local magnified images of $S1$ and $S2$.} 
        \label{A2}
    \end{figure}

 To more clearly present the polarization distribution of the ultra-high-order Stokes skyrmions we generated, Fig.\ref{A1} and Fig.\ref{A2} show the Stokes parameters of the vector beams obtained by coherent superposition of LG and PV beams, respectively.

Since the Stokes parameter $S0$ represents the complex amplitude distribution in the beam cross-section, it is evident why we refer to the generated ultra-high-order skyrmion as ring-shaped. The Stokes parameters $S1$ and $S2$ represent the distributions of horizontal and diagonal polarizations, respectively. For a skyrmion, they should show $2N$ alternating red and blue stripes along the azimuthal direction, meaning there are 800 stripes in Fig.\ref{A1} and 200 stripes in Fig.\ref{A2}. Due to the limited number of pixels occupied by $S1$ and $S2$ within a single azimuthal period, only Moiré fringes are visible in the overall view. Therefore, local magnified images are provided. The results of the magnified images show that $S1$ and $S2$ differ by a $\pi/4$ rotation angle. Combined with precise numerical integration methods, we obtained the better ultra-high-order topological numbers presented in the main text.


\end{document}